\documentclass[lettersize,journal]{IEEEtran}
\usepackage{amsmath,amssymb,amsfonts,amsthm}
\usepackage{algorithmic}
\usepackage{array}
\usepackage{textcomp}
\usepackage{stfloats}
\usepackage{url}
\usepackage{verbatim}
\usepackage{graphicx}
\usepackage{subcaption}
\usepackage{epstopdf}
\usepackage{cite}
\usepackage{xcolor}
\usepackage{diagbox}
\usepackage{slashbox}
\usepackage{siunitx}
\usepackage[colorlinks,
            linkcolor=blue,     
            anchorcolor=blue, 
            citecolor=blue,       
            ]{hyperref}
\usepackage{multirow}
\usepackage{tikz}
\usepackage{tikz-3dplot}
\usepackage{svg}
\usepackage{overpic}
\usetikzlibrary {arrows.meta,bending}

\usepackage[linesnumbered,lined,boxed,commentsnumbered,ruled,vlined]{algorithm2e}

\usepackage{pgfplots}
\pgfplotsset{compat=newest}

\usepackage{xifthen}
\newcommand{\prob}[1][]{
\ifthenelse{\isempty{#1}}%
      {\ensuremath{P}}%
    {\ensuremath{P\left\(#1\right\)}}%
}
\newcommand{\vect}[1]{\boldsymbol{\mathrm{#1}}}

\usepackage[acronym]{glossaries}
\makeglossaries
\glsdisablehyper
\newacronym{ml}{ML}{maximum likelihood}
\newacronym{gps}{GPs}{Gaussian processes}
\newacronym{gp}{GP}{Gaussian process}
\newacronym{mlp}{MLP}{multilayer perceptrons}
\newacronym{led}{LED}{light-emitting diode}
\newacronym{mse}{MSE}{mean squared error}
\newacronym{crlb}{CRLB}{Cramer-Rao lower bound}
\newacronym{fim}{FIM}{Fisher Information matrix}
\newacronym{rss}{RSS}{received signal strength}
\newacronym{se}{SE}{squared error}
\newacronym{snr}{SNR}{Signal-to-Noise Ratio}
\newacronym{cdf}{CDF}{cumulative distribution function}
\newacronym{ls}{LS}{least squares}
\newacronym{toa}{TOA}{time-of-arrival}
\newacronym{tdoa}{TDOA}{time-difference-of-arrival}
\newacronym{aoa}{AOA}{angle of arrival}
\newacronym{uwb}{UWB}{ultra-wideband}
\newacronym{nlos}{NLOS}{non-line-of-sight}
\newacronym{los}{LOS}{line-of-sight}
\newacronym{svm}{SVM}{support vector machines}
\newacronym{sdps}{SDPs}{semidefinite programs}
\newacronym{map}{MAP}{maximum a posteriori}
\newacronym{owp}{OWP}{optical wireless positioning}
\newacronym{imu}{IMU}{inertial measurement unit}
\newacronym{gt}{GT}{ground truth}
\newacronym{pd}{PD}{photodiode}
\newacronym{ekf}{EKF}{Extended Kalman Filter}
\newacronym{zupt}{ZUPT}{Zero velocity UPdaTe}
\newacronym{rbls}{RBLS}{range-based localization service}
\newacronym{ins}{INS}{inertial navigation system}

\usepackage{pifont} 
\usepackage{multirow}
\usepackage{arydshln}
\usepackage{booktabs} 
\usepackage{url}
    
\begin{document}

\title{OWP-IMU: An RSS-based Optical Wireless  and IMU Indoor Positioning Dataset}



\author{Fan Wu, Jorik De Bruycker, Daan Delabie, Nobby Stevens~\IEEEmembership{Member,~IEEE,} François Rottenberg~\IEEEmembership{Member,~IEEE,} and  Lieven De Strycker~\IEEEmembership{Member,~IEEE,}
\thanks{The authors are with \href{https://iiw.kuleuven.be/onderzoek/dramco}{ESAT-DRAMCO}, KU Leuven (Ghent), 9000 Ghent, Belgium (e-mail: fan.wu@kuleuven.be).}

\thanks{ Acknowledge the Chinese Scholarship Council (CSC) for the Ph.D
grant of Fan Wu (No. 202106340043)
This work has been submitted to the IEEE for possible publication. Copyright may be transferred without notice, after which this version may no longer be accessible.
}}

\markboth{Journal of \LaTeX\ Class Files,~Vol.~14, No.~8, August~2021}%
{Shell \MakeLowercase{\textit{et al.}}: A Sample Article Using IEEEtran.cls for IEEE Journals}


\maketitle

\begin{abstract}
\Gls{rss}-based \gls{owp} systems are becoming popular for indoor localization because they are low-cost and accurate. However, few open-source datasets are available to test and analyze RSS-based OWP systems. In this paper, we collected RSS values at a sampling frequency of \SI{27}{\Hz}, \gls{imu} at a sampling frequency of \SI{200}{\Hz} and the ground truth at a sampling frequency of \SI{160}{\Hz} in two indoor environments. One environment has no obstacles, and the other has a metal column as an obstacle to represent a \gls{nlos} scenario. We recorded data with a vehicle at three different speeds (low, medium and high). The dataset includes over \SI{110}{k} data points and covers more than \SI{80}{\min}. We also provide benchmark tests to show localization performance using only RSS-based OWP and improve accuracy by combining IMU data via extended kalman filter. The dataset OWP-IMU is open source\footnote{\url{https://dramco.github.io/OWP-IMU-Dataset/}} to support further research on indoor localization methods.
\end{abstract}

\begin{IEEEkeywords}
received signal strength, visible light positioning, optical wireless positioning, inertial measurement unit, indoor localization.
\end{IEEEkeywords}

\section{Introduction}
\IEEEPARstart{R}{ange}-based localization services (RBLSs) have been adopted in many applications, such as airports navigation, warehouse management \cite{zhu2024survey}. In terms of  various and distinct business requirements, the appropriate choice of  the physical signal and the distance measurement method as the main components of RBLS is necessary. Thanks to several years of research and development, various combinations of  physical signals with distance measure methods  enriched the applicability of RBLS.  The physical signal contains \gls{uwb}~\cite{wymeersch2012machine}, WiFi~\cite{9918126} and \gls{owp} (e.g., visible light, infrared light)~\cite{10902492}, and distance measure methods consist of \gls{rss}~\cite{8675356} , \gls{toa}~\cite{10290120} and \gls{tdoa}~\cite{10592650}. Among them,  \gls{rss}-based \gls{owp} is performing  well since its cost-efficiency and easy to  install. However, it's still challenged by multipath effects~\cite{1633354,864015} and \gls{nlos} contributions~\cite{4290044,guvencc2007nlos}.

Recently, many studies~\cite{sun2025tightly,8768057,cheng2020single,8911760} have proposed combining \gls{imu} with \gls{owp} to improve localization robustness, especially  under multipath  and \gls{nlos} conditions. However, most existing papers only present the methods and results without providing open datasets. Tab.~\ref{tab:open-source data} summarizes the open source datasets related to \gls{owp}. For instance, the dataset from~\cite{f28n-6292-20} contains 158 measurement points for visible light positioning  without incorporating \gls{imu} data. These points also do  not  represent continuous trajectory, and the provided \gls{gt}  includes only  coordinates without orientation. Another dataset from~\cite{9zsx-bm77-23} uses a camera to measure visible light \gls{rss} and estimate both coordinates and orientation based on \gls{aoa}. However, it includes only 98 discrete measurements points.

\renewcommand{\arraystretch}{1}
\begin{table}[!h]
  \centering
  \caption{Overview of Public Optical Wireless Positioning RSS-based Datasets.}
  \label{tab:open-source data}
  \begin{tabular}{cccc}
  \hline
  \diagbox{\textbf{Attribute}}{\textbf{Dataset}}& \begin{tabular}[c]{@{}c@{}}RSS-based\\ OWP\cite{f28n-6292-20}\end{tabular} & \begin{tabular}[c]{@{}c@{}}AOA-based\\ OWP\cite{9zsx-bm77-23}\end{tabular} & \begin{tabular}[c]{@{}c@{}}\textbf{OWP-IMU}\\ \textbf{Ours}\end{tabular} \\ \hline
  Receiver Type & PD & Camera & PD \\
  LOS/NLOS Testing & \ding{55} & \ding{55} & $\checkmark$ \\
  IMU Integration & \ding{55} & \ding{55} & $\checkmark$ \\
  GT Orientation & \ding{55} & $\checkmark$ & $\checkmark$ \\
  Continuous Trajectory & \ding{55} & \ding{55} & $\checkmark$ \\
  Receiver Total Captures & 158 & 98 & $>$ \SI{110}{k} \\
  Open Access & \ding{55} & $\checkmark$ & $\checkmark$ \\
  \hline
  \end{tabular}
\end{table}
\renewcommand{\arraystretch}{1}

To foster further research in this domain, this paper introduces a comprehensive RSS-based \gls{owp} and \gls{imu} indoor localization dataset. We collected data in two distinct environments to examine the impact of obstacles on positioning performance. One scenario is an  obstacle-free,  and the other contains a metal column obstacle. In both scenarios, a vehicle equipped with an \gls{imu}, \gls{pd}, and \gls{gt} markers move along continuous trajectories at three different velocities: low (\SI{0.15}{m/s}), medium (\SI{0.275}{m/s}), and high (\SI{0.45}{m/s}). For each trajectory, we recorded \gls{gt} data (position and orientation), RSS values from infrared lights received by a \gls{pd}, and \gls{imu} readings. As illustrated  in Tab.~\ref{tab:open-source data}, the dataset comprises more than \SI{110}{k}  sampled points, covering over  \SI{80}{\min} of data collection. 

Furthermore, we provide benchmark tests using our dataset to demonstrate its usability. Firstly, we implemented \gls{gps}  regression relying solely on the \gls{owp} system,  following approaches commonly adopted in papers \cite{9079878,zeng2021data}. Secondly, we applied an \gls{ekf} to fuse the \gls{imu} and \gls{owp}  to enhancing positioning accuracy and robustness.

The main contributions of this paper are as follows:
\begin{enumerate}
    \item To the best of our knowledge, this is the first open-source dataset  combining RSS-based \gls{owp}, \gls{imu}, and accurate \gls{gt} data. It has  measurements taken at high sampling rates: \gls{owp} data at \SI{27}{Hz}, \gls{imu} readings at \SI{200}{Hz}, and \gls{gt} at \SI{160}{Hz}. It also has trajectories recorded at three different velocities under both \gls{los} and \gls{nlos} conditions together over \SI{80}{\min}. 
    
    \item  We offer full benchmarks to show how to use the dataset. They include localization tests (\gls{gps}, multilateration~\cite{zhuang2018survey}) using only the RSS-based \gls{owp} system  and performance gains from sensor fusion with an \gls{ekf}. From these tests, using an \gls{ekf} with \gls{imu} lowers the P99 localization error from about \SI{40}{cm} to about \SI{25}{cm}, a \SI{45}{\percent} reduction.
\end{enumerate}

The paper is organized as follows. In Section~\ref{sec:SYSTEM DESCRIPTION}, our system is described. Section~\ref{sec:Data Collection} presents the datasets that we have collected. Then in Section~\ref{sec:Benchmarks}, we use some benchmarks to evaluate and validate our datasets. Finally, Section~\ref{sec:CONCLUSION} concludes the paper.

\section{RSS-based Optical Wireless and IMU Positioning System}
\label{sec:SYSTEM DESCRIPTION}

\begin{figure}[!h]
  \centering
  \includegraphics[width=\linewidth]{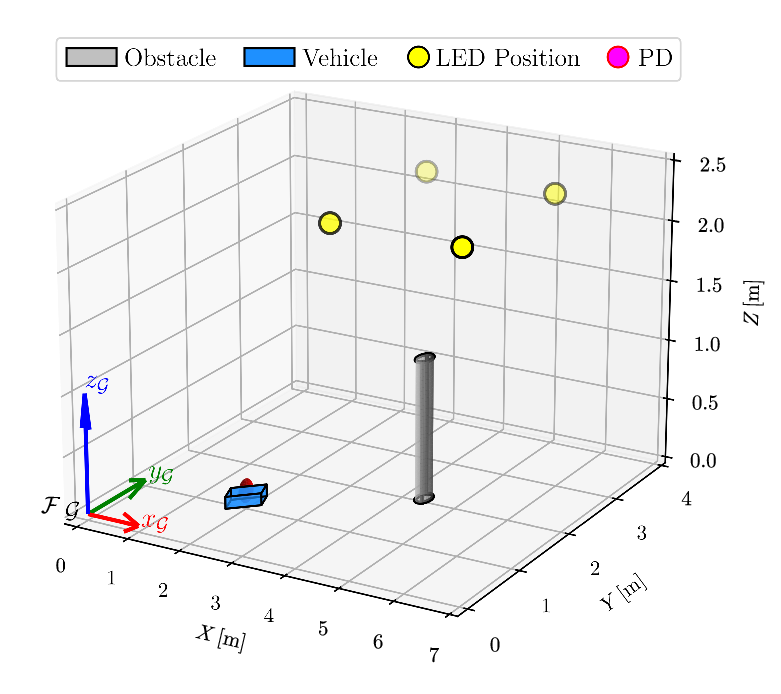}
  \caption{Illustration of the experimental setup for data collection.}
  \label{fig:setup_sketch}
\end{figure}

Our RSS-based \gls{owp} and \gls{imu} Positioning System is illustrated in Fig.~\ref{fig:setup_sketch}. In this figure, 
four LEDs  (yellow circles) are installed on the ceiling at the same height. The origin of the  global frame $\mathcal{F}_{\mathcal{G}}$ is  at one corner of the experimental area. The axes follow the right-hand rule. The Ackermann \cite{wikipedia:ackermann_steering}  vehicle (blue cuboid) carries a \gls{pd} receiver (pink circle) on its top surface. In order to investigate localization performance under different conditions, we conducted experiments in two scenarios: one without obstacle and another including a metal column obstacle. The obstacle (grey cylinder) is placed at a fixed position. The detailed setup parameters are summarized in Tab.~\ref{tab:setup}  and include the positions and dimensions of LEDs, the \gls{pd} height, and obstacle specifications. 

\renewcommand{\arraystretch}{1.1}
\begin{table}[!h]
  \centering
  \caption{Platform Setup Parameters}
  \label{tab:setup}
  \begin{tabular}{l c}
  \hline
  \textbf{Parameters} & \textbf{Values} \\
  \hline
  Ceiling Height from Ground & \SI{2.4}{\meter} \\
  PD Height from Ground & \SI{20}{\centi\meter} \\
  Number of LEDs & 4 \\
  LED Positions [m] & [\SI{3.561}{}, \SI{1.080}{}], [\SI{3.561}{}, \SI{2.910}{}] \\
                   & [\SI{5.975}{}, \SI{1.080}{}], [\SI{5.975}{}, \SI{2.910}{}] \\
  Obstacle Diameter & $\phi=$~\SI{24.5}{\centi\meter} \\
  Obstacle Height & \SI{1.16}{\meter} \\
  Obstacle Position [m] & [\SI{4.471}{}, \SI{1.947}{}] \\
  \hline
  \end{tabular}
\end{table}

\begin{figure}[!h]
    \centering
    \includegraphics[width=0.8\linewidth]{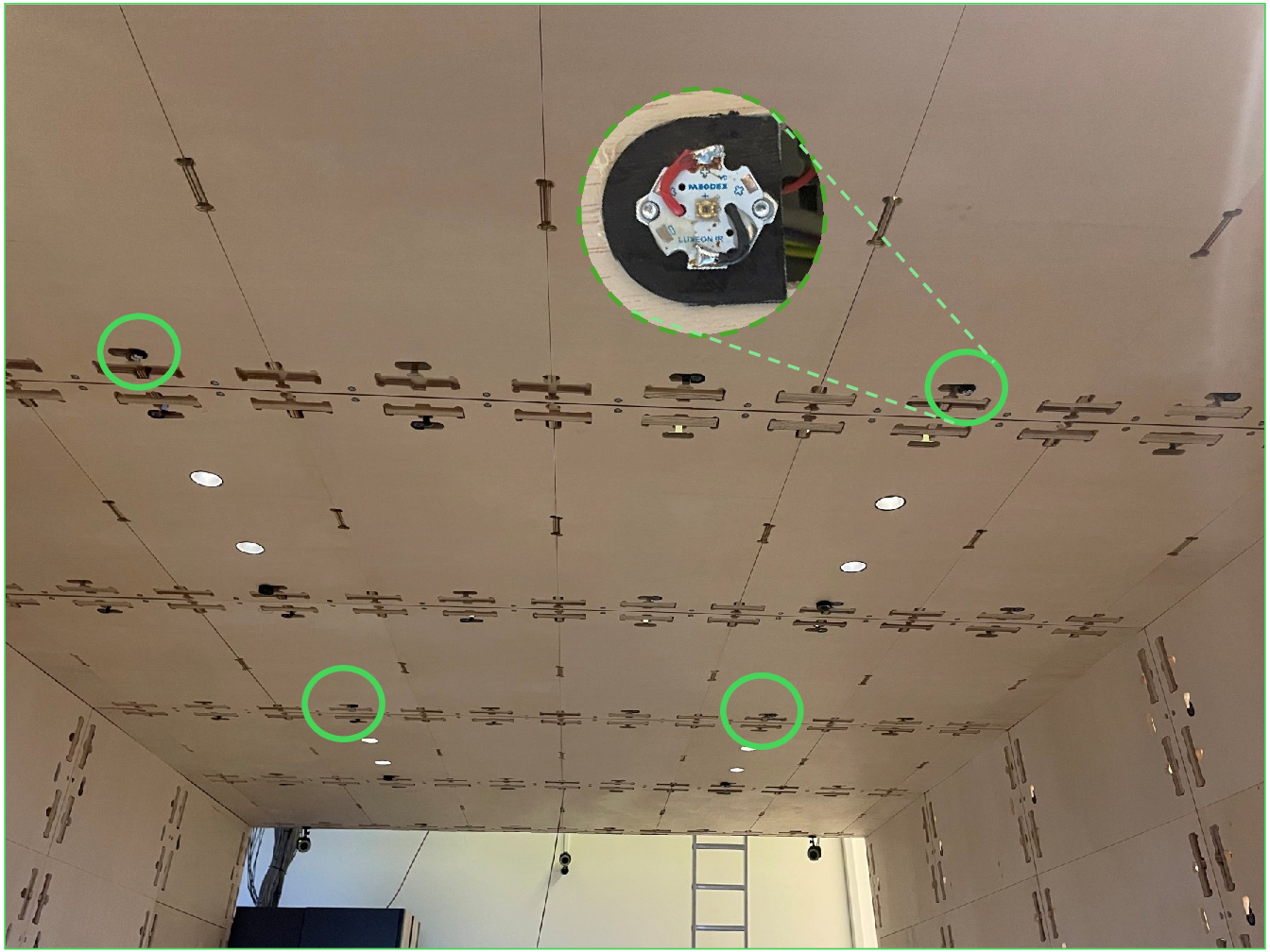}
    \caption{Our laboratory setup of the infrared LED transmitters for optical wireless localization.}
    \label{fig:ceiling}
\end{figure}

Fig.~\ref{fig:ceiling} shows the actual setup constructed in our laboratory. Several infrared LEDs \cite{luxeon_ir_domed_line} (marked with green circles) are mounted on the ceiling. Each LED emits infrared light modulated at a unique frequency. This light is received by the \gls{pd} installed on the vehicle. The \gls{rss} at the \gls{pd} depends on various factors, including the distance between the LED and the receiver, transmitter and receiver tilt angles, the Lambertian emission characteristics, multipath, and \gls{nlos} conditions \cite{10457033}. To effectively study and analyze these factors, we collected data with the vehicle moving at different velocities and under scenarios with and without an obstacle.

\begin{figure}[]
    \centering
    \includegraphics[width=\linewidth]{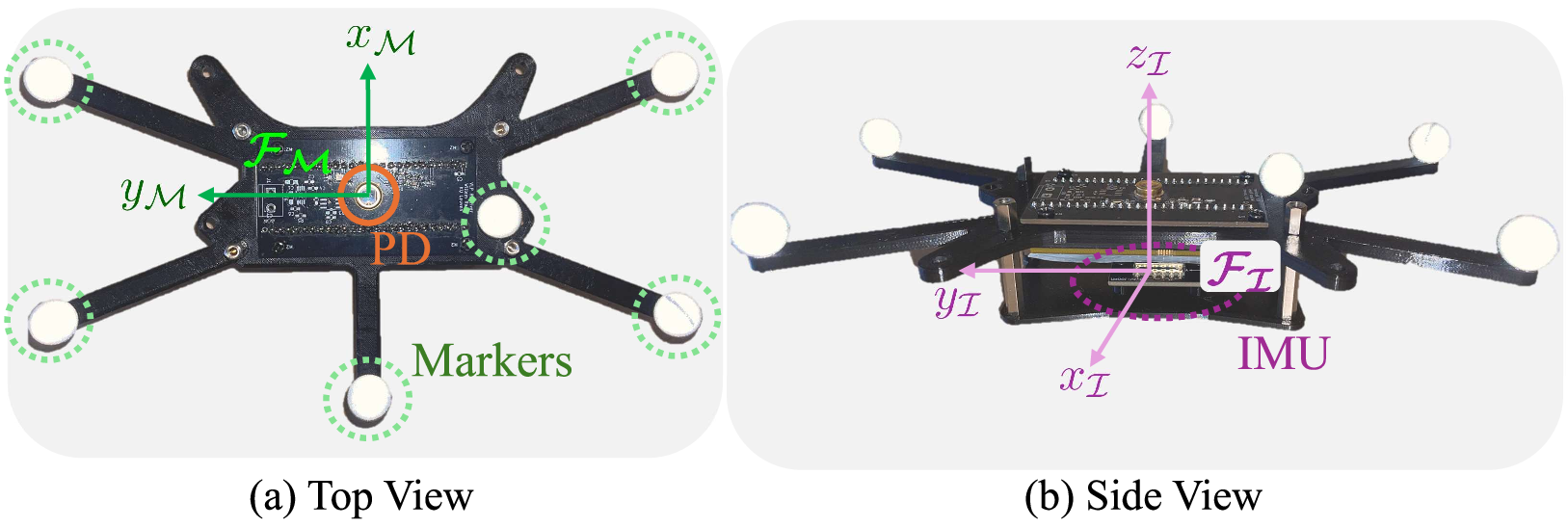}
    \caption{Top and side views of our customized platform combining ground truth markers, an IMU, and a PD receiver.}
    \label{fig:IMU_PD_device}
\end{figure}

We designed a customized platform  combining the \gls{gt} markers, \gls{pd} and \gls{imu}.  Fig.~\ref{fig:IMU_PD_device} illustrates the top and side views of our platform. In the top view, the markers used by the Qualisys motion capture system \cite{qualisys_miqus} are highlighted with green dashed circles. Qualisys identifies these markers and returns the local coordinate frame $\mathcal{F}_\mathcal{M}$. In this local frame, the $x_\mathcal{M}$ axis points forward along the vehicle's moving direction, the $y_\mathcal{M}$ axis points laterally, and the $z_\mathcal{M}$ axis aligns vertically, parallel to the global frame $\mathcal{F}_\mathcal{G}$. The origin of $\mathcal{F}_\mathcal{M}$ coincides with the location of the \gls{pd}, indicated by an orange circle.

The side view shows the \gls{imu} is installed directly under the \gls{pd} at a distance of \SI{19}{mm}. During installation, we aligned the local frame  $\mathcal{F}_\mathcal{I}$ of the \gls{imu} to be parallel with the frame  $\mathcal{F}_\mathcal{M}$. 

\section{Data Collection}
\label{sec:Data Collection}
\begin{figure*}[]
  \centering
  \includegraphics[width=\linewidth]{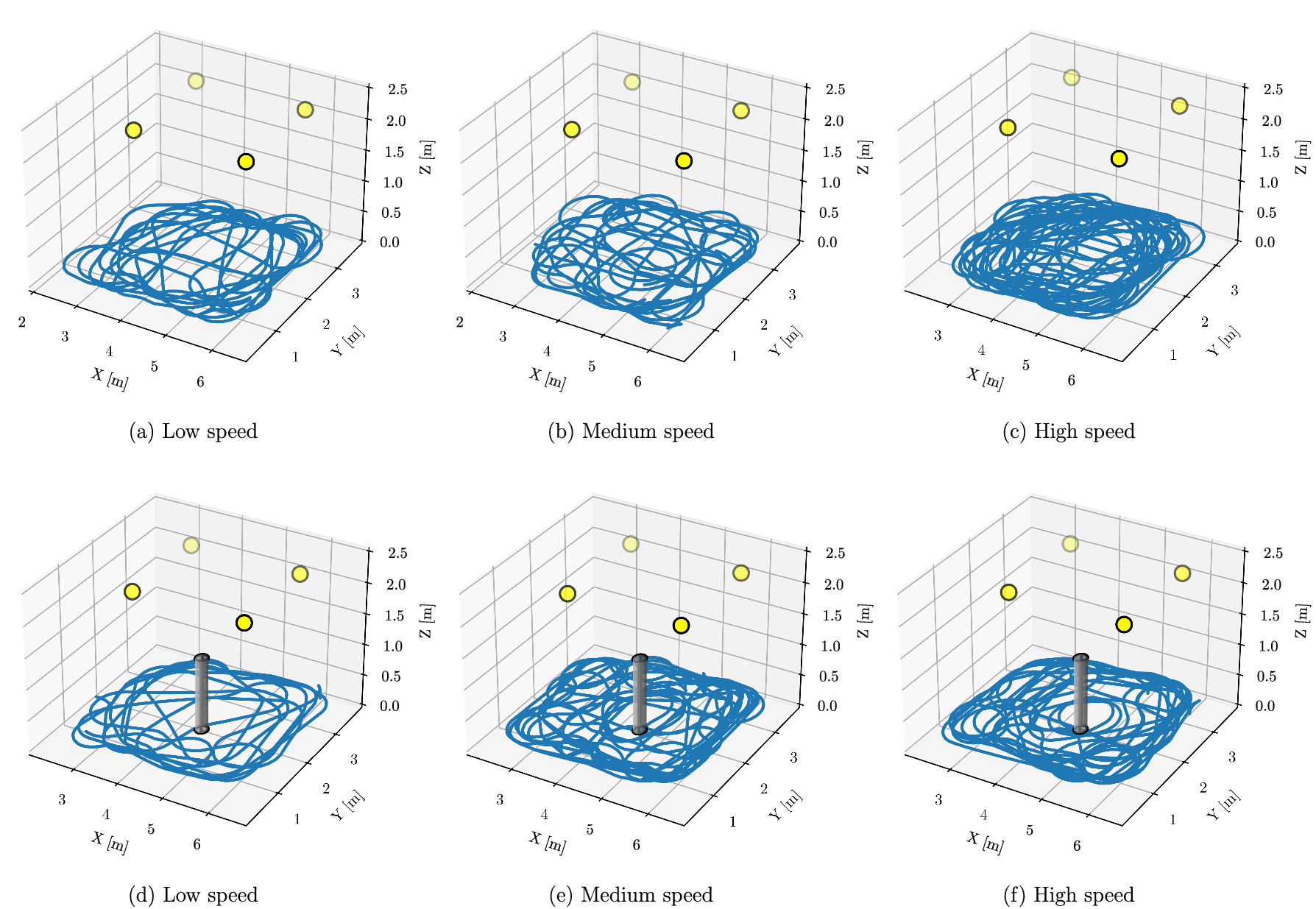}
  \caption{Trajectories recorded by the vehicle at different speeds with or without the obstacle.}
  \label{fig:trajectory}
\end{figure*}

To evaluate the robustness of our localization system, we collected datasets under three different velocities:  low speed (\SI{0.15}{m/s}), medium speed (\SI{0.275}{m/s}) and high speed (\SI{0.45}{m/s}). Fig.~\ref{fig:trajectory} shows trajectories derived from \gls{gt}. The first row represents the data in the obstacle-free environment. The bottom row shows trajectories in the presence of the metal column obstacle. Each trajectory represents a  recording  of at least \SI{10}{min}. As expected, the trajectories at  higher speeds (rightmost columns) cover longer distances. 

\begin{figure}[]
  \centering
  \includegraphics[width=\linewidth]{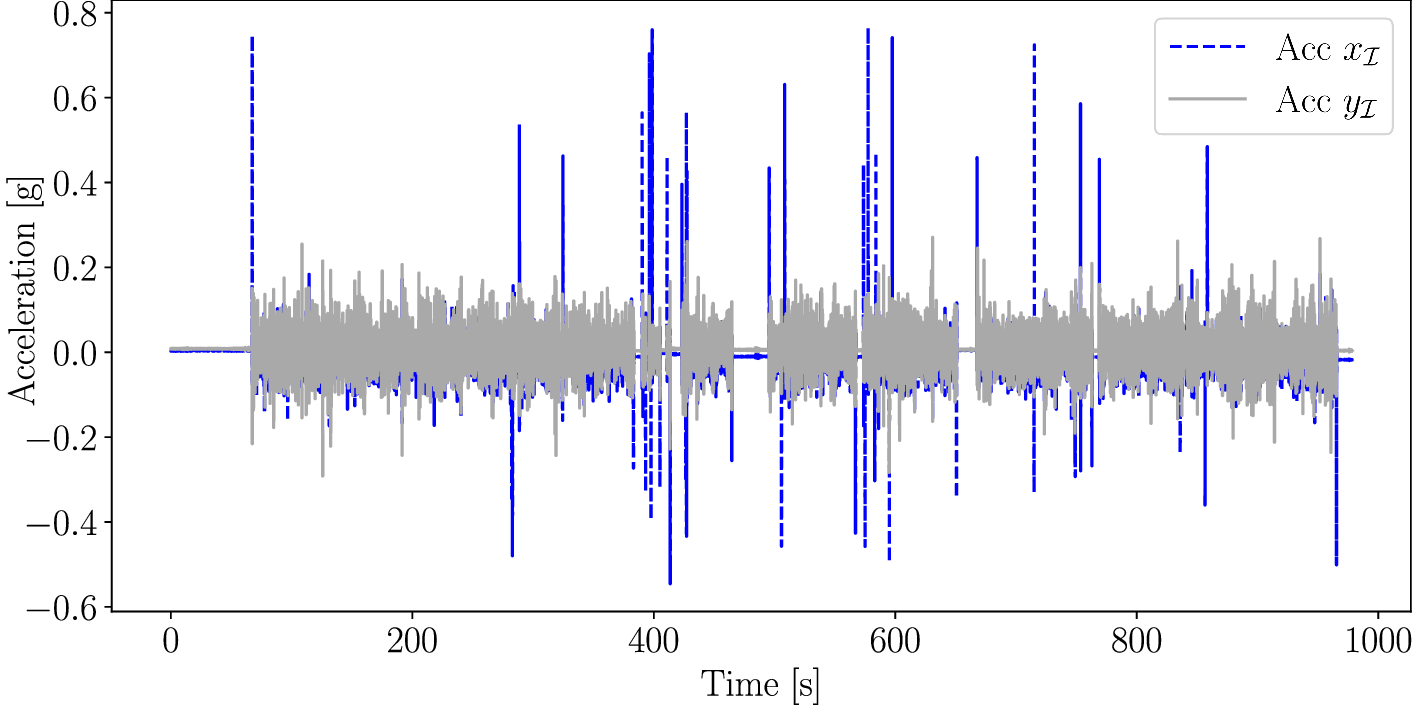}
  \caption{IMU acceleration measurement example under low speed and obstacle-free conditions. }
  \label{fig:imu_display_acc}
\end{figure}

The vehicle utilizes Ackermann steering geometry. Under ideal conditions, it does not experience lateral slipping. The \gls{imu} is aligned with the vehicle, with $x_\mathcal{I}$ pointing forward direction, and $y_\mathcal{I}$ axis pointing lateral direction. Fig.~\ref{fig:imu_display_acc} shows the measured accelerations in the $x_\mathcal{I}$ and $y_\mathcal{I}$ directions. The measured accelerations reflect typical vehicle behaviors, such as acceleration, deceleration, turning, and stopping. Notably, periods when both acceleration signals ($x_\mathcal{I}$ and $y_\mathcal{I}$) remain close to zero indicate stationary conditions. This stationary signal is valuable for implementing \gls{zupt} \cite{ramanandan2011inertial} method, 
which is used to correct drift in \gls{ins}. By detecting when the vehicle is motionless, the estimated velocity from the \gls{imu} can be reset to zero, thereby limiting the accumulation of integration errors.
To facilitate performance analysis, the vehicle was kept stationary for approximately \SI{1}{min} at the beginning of each recording. Finally, we got three datasets GT, IMU and OWP for each trajectory as Tab.~\ref{tab:datasets format} shows.

\renewcommand{\arraystretch}{1.2}
\begin{table}[]
  \centering
  \caption{Recorded datasets format}
  \label{tab:datasets format}
  \begin{tabular}{ccl}
  \hline
  \begin{tabular}[c]{@{}c@{}}Datasets\end{tabular} &
    \begin{tabular}[c]{@{}c@{}}CSV \\ Column\end{tabular} &
    Format \\ \hline
  \multicolumn{1}{c}{\multirow{3}{*}{GT}}  & 0        & Timestamp {[}s{]}                                                   \\ 
  \multicolumn{1}{c}{}                     & 1$\sim$3 & Coordinates $(x_\mathcal{M},y_\mathcal{M},z_\mathcal{M})$ {[}m{]}   \\ 
  \multicolumn{1}{c}{} &
    4$\sim$12 &
    \begin{tabular}[c]{@{}l@{}}Rotation Matrix (3x3)\\ x-y-z rotation order, clockwise positive\end{tabular} \\ \hline\hline

  \multicolumn{1}{c}{\multirow{3}{*}{IMU}} & 0        & Timestamp {[}s{]}                                                   \\ 
  \multicolumn{1}{c}{}                     & 1$\sim$3 & Acceleration $(x_\mathcal{I},y_\mathcal{I},z_\mathcal{I})$ {[}g=\SI{9.81}{m/s^2}{]}  \\ 
  \multicolumn{1}{c}{} &
    4$\sim$6 &
    \begin{tabular}[c]{@{}l@{}}Gyroscope $(x_\mathcal{I},y_\mathcal{I},z_\mathcal{I})$ {[}deg/s{]}\\ x-y-z rotation order, anticlockwise positive\end{tabular} \\ \hline\hline
    
  \multicolumn{1}{c}{\multirow{2}{*}{OWP}}  & 0        & Timestamp {[}s{]}                                                   \\ 
  \multicolumn{1}{c}{}                     & 1$\sim$4 & RSS values ($\text{RSS}_1, \text{RSS}_2,\text{RSS}_3,\text{RSS}_4$) \\ \hline
  \end{tabular}
  \end{table}

\section{Benchmarks}
\label{sec:Benchmarks}
In this section, firstly, we use the machine learning \gls{gps} and traditional method multilateration to regress the \gls{rss} values solely based on \gls{owp}  system. Secondly, we adopted \gls{ekf} to fuse the \gls{imu} with the \gls{owp} to achieve better estimation accuracy.

\subsection{OWP Benchmarks}
Since the OWP sampling rate \SI{27}{Hz} is much lower than the GT sampling rate \SI{160}{Hz}, for each OWP timestamp we find the closest GT timestamp and take its coordinates as the true position for that sample. The total number of synced datasets is about 17000. Then we applied \gls{gps} regression and multilateration. In \gls{gps}, the four-dimensional RSS vector serves as input and the 2-D position $(x,y)$ as output. We train the model on a random subset of data and test it on the remaining data. We repeat this split 10 times and report P50 and P99 errors.

\begin{figure*}[]
    \centering
    \begin{subfigure}[b]{0.335\textwidth}
        \centering
        \includegraphics[width=\textwidth]{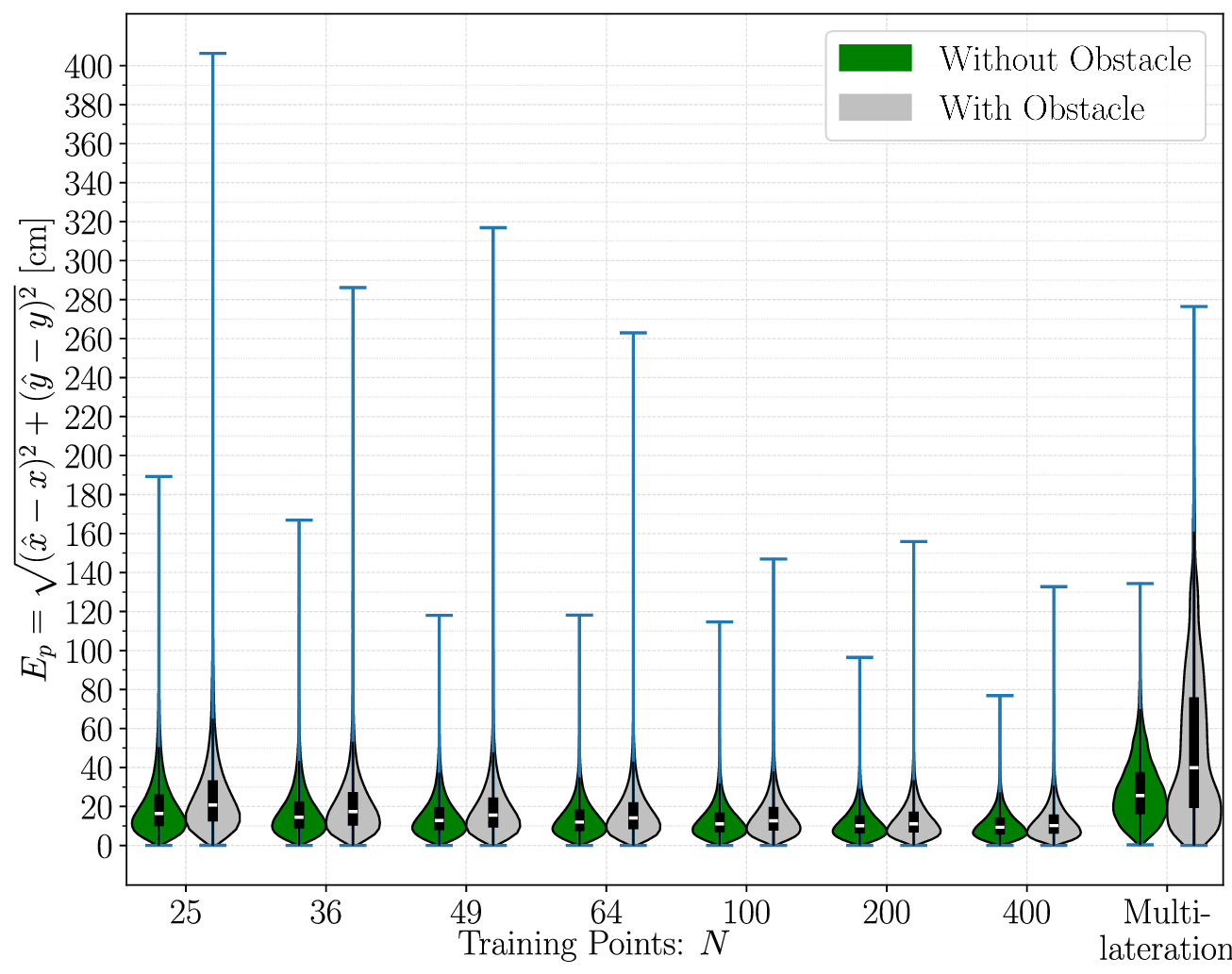}
        \caption{Low speed}
        \label{fig:sub1}
    \end{subfigure}
    \hfill
    \begin{subfigure}[b]{0.32\textwidth}
        \centering
        \includegraphics[width=\textwidth]{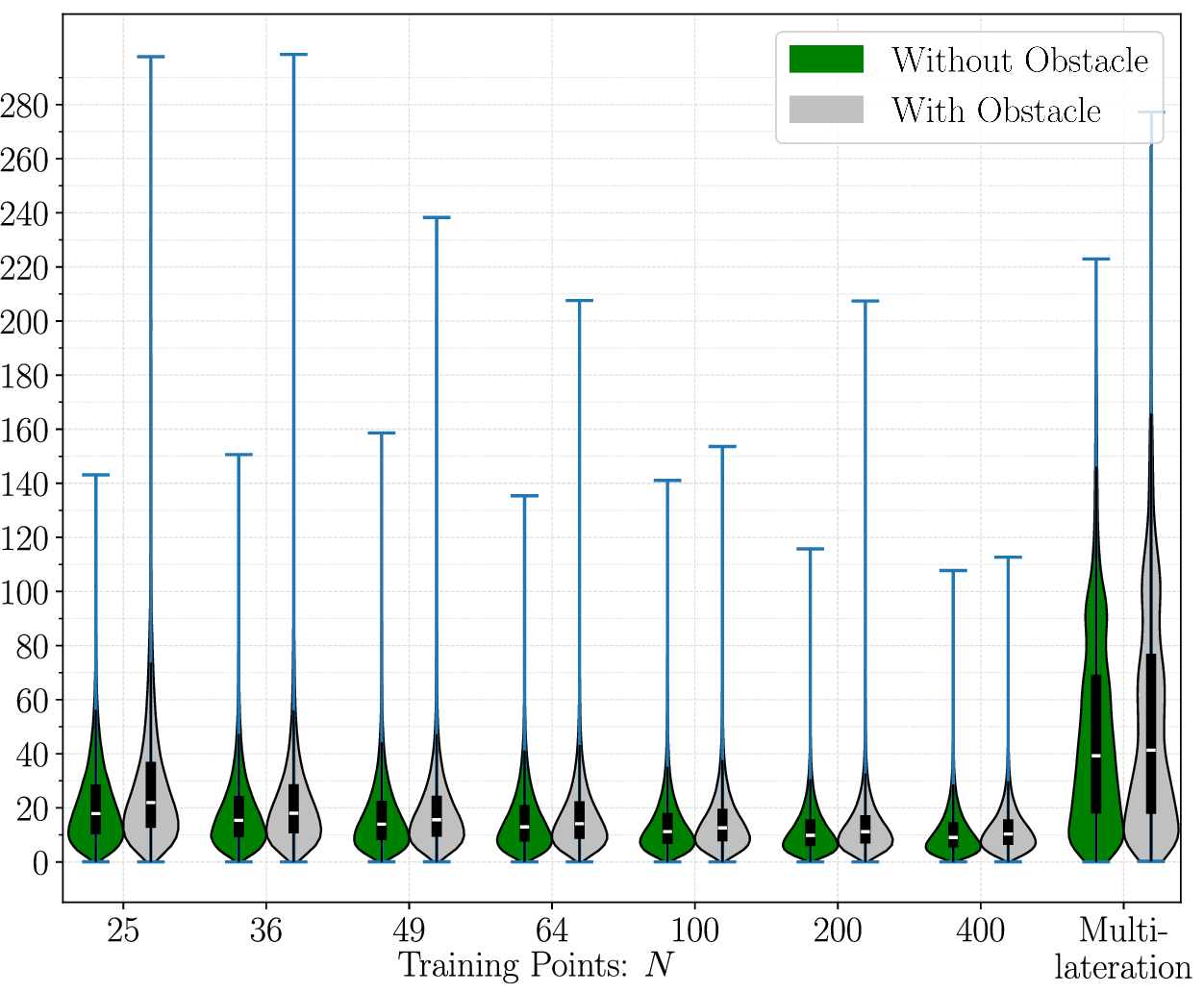}
        \caption{Medium speed}
        \label{fig:sub2}
    \end{subfigure}
    \hfill
    \begin{subfigure}[b]{0.32\textwidth}
        \centering
        \includegraphics[width=\textwidth]{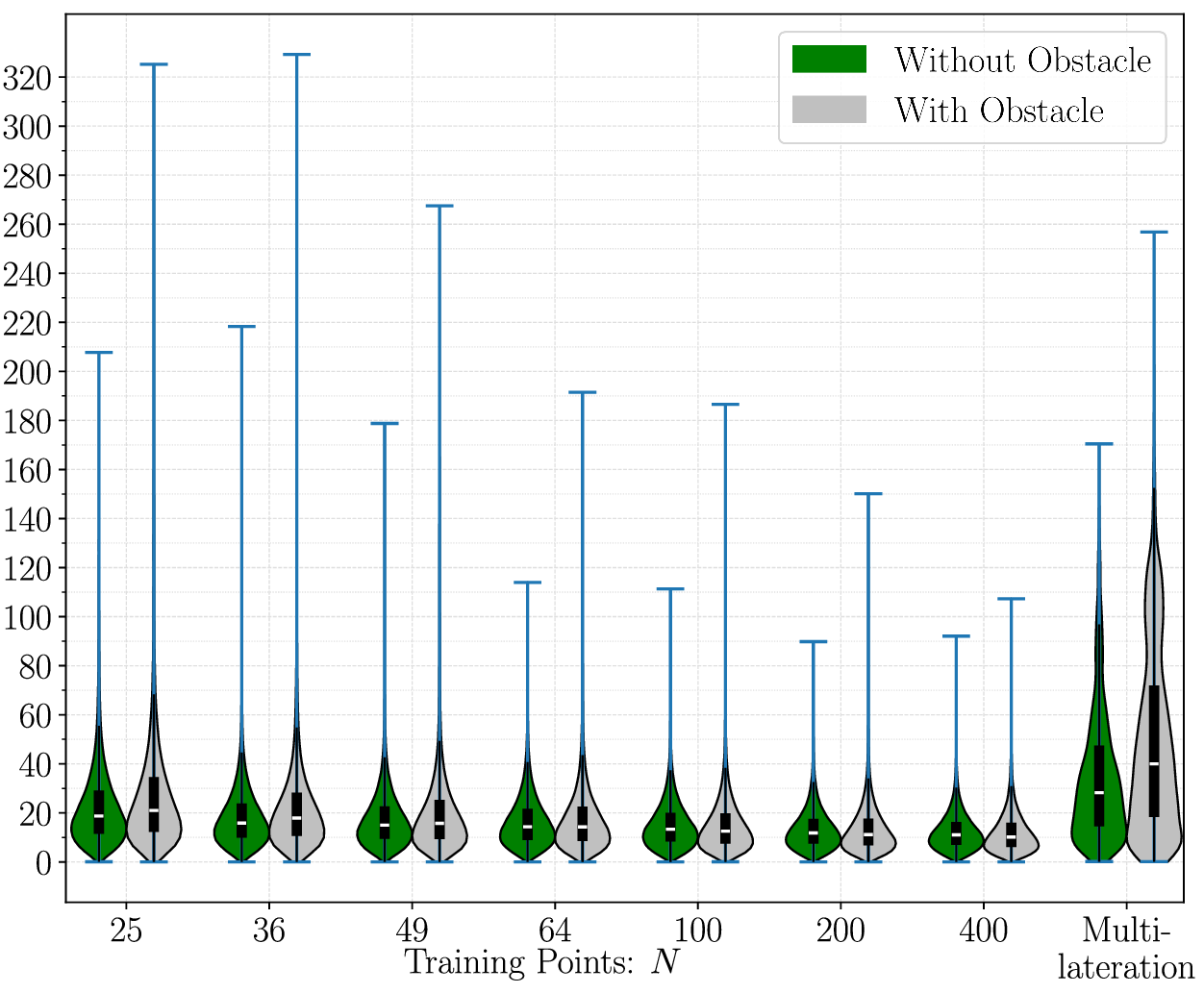}
        \caption{High speed}
        \label{fig:sub3}
    \end{subfigure}
    \caption{Error distributions of \gls{gps} training models at three vehicle speeds. The green plots represent the obstacle-free scenario, while the gray plots represent the obstacle-present scenario.}
    \label{fig:GPs_results}
\end{figure*}

Multilateration assumes no tilt in the LEDs or the PD. Under this assumption, the RSS value is proportional to $1/d^{m+3}$, where $d$ is the optical wireless propagation distance and $m$ is the Lambertian order. With the infrared LEDs we deployed, $m=0.5$, so the exponent is $3.5$. We first calibrate a gain factor for each LED by all measurement datasets. Then we use it  to compute each LED to PD distance from its RSS reading. Finally, we solve for the receiver position by a least-squares intersection of the four distance estimates. Fig.~\ref{fig:GPs_results} shows the resulting error distributions, and Tab.~\ref{tab:gp results} lists the P50 and P99 percentile errors. Based on these results, we found

\begin{itemize}
  \item \textbf{GPs vs.\ multilateration}: Even with just 16 training samples, GP regression achieves a P50 error of around \SI{20}{cm}, outperforming multilateration’s~\SI{27}{cm}. With 400 training samples, GP’s P50 drops to~\SI{10}{cm}, while multilateration remains near \SI{25}{cm}. This demonstrates GP’s superior accuracy across all sample sizes.

  \item \textbf{Obstacle impact}: The obstacle consistently causes larger errors than the obstacle-free setup for both methods GPs and multilateration. For GPs, this gap is more apparent when the training set is small (e.g., 16 or 25 samples). This result implies that obstacles introduce additional complexity that is harder to capture with limited training data. The quality of the multilateration estimation also decays significantly. Shadows and multipath create outliers in its distance estimates.
  
  \item \textbf{Training size impact}: Increasing the number of training samples  improves GPs accuracy. With 400 training points, the P50 error approaches \SI{10}{cm}, regardless of speed or obstacle presence.
  
  \item \textbf{Speed effect}: Vehicle speed has minimal impact on OWP accuracy. With only 16 samples, GP regression yields roughly \SI{20}{\cm} P50 errors across low, medium, and high speeds. At 400 samples, all speeds converge to about \SI{10}{\cm} (P50) and \SI{36}{\cm} (P99). However, at higher speeds the fixed \SI{27}{Hz} sampling produces fewer trajectory points, which can reduce the smoothness and continuity of the estimated path.
 
\end{itemize}

\begin{table}[]
  \centering
  \caption{Localization P50/P99 errors [cm] for GPs vs. multilateration using only OWP data.}
  \label{tab:gp results}
  \begin{tabular}{cccccc}
    \toprule
    \multirow{2}{*}{Speed} & \multirow{2}{*}{\shortstack{Training Size $N$\\ Or Multilateration}} & \multicolumn{2}{c}{Without Obstacle} & \multicolumn{2}{c}{With Obstacle} \\
    \cmidrule(lr){3-4} \cmidrule(lr){5-6}
     & & P50 & P99 & P50 & P99 \\
    \midrule
    \multirow{8}{*}{\shortstack{Low \\ \SI{0.15}{m/s}}} & 16 & 18.79 & 87.96 & 24.16 & 164.12 \\
      & 25 & 16.40 & 79.31 & 20.78 & 147.33 \\
      & 36 & 14.49 & 60.98 & 17.37 & 115.68 \\
      & 49 & 12.76 & 51.94 & 15.56 & 94.43 \\
      & 64 & 12.02 & 47.92 & 14.09 & 94.91 \\
      & 100 & 11.08 & 42.41 & 12.68 & 60.29 \\
      & 200 & 10.11 & 38.37 & 11.09 & 49.79 \\
      & 400 & 9.33 & 34.75 & 10.21 & 42.13 \\
      \cdashline{2-6}
      & Multilateration & 25.59 & 84.16 & 39.88 & 167.56 \\
    \midrule
    \multirow{8}{*}{\shortstack{Medium \\ \SI{0.275}{m/s}}} & 16 & 23.43 & 110.25 & 25.89 & 190.04 \\
      & 25 & 17.89 & 73.07 & 21.96 & 157.34 \\
      & 36 & 15.41 & 66.39 & 18.02 & 147.77 \\
      & 49 & 13.99 & 62.14 & 15.60 & 92.11 \\
      & 64 & 12.95 & 57.05 & 14.16 & 80.44 \\
      & 100 & 11.21 & 49.98 & 12.58 & 55.21 \\
      & 200 & 9.83 & 41.86 & 11.17 & 47.91 \\
      & 400 & 9.05 & 38.85 & 10.33 & 40.42 \\
      \cdashline{2-6}
      & Multilateration &  39.26 & 155.08 & 41.34 & 159.18 \\
    \midrule
    \multirow{8}{*}{\shortstack{High \\ \SI{0.45}{m/s}}} & 16 & 22.50 & 104.05 & 26.16 & 200.99 \\
      & 25 & 18.73 & 82.19 & 21.00 & 151.46 \\
      & 36 & 15.84 & 59.53 & 17.89 & 142.93 \\
      & 49 & 15.03 & 57.05 & 15.77 & 116.01 \\
      & 64 & 14.35 & 51.07 & 14.30 & 88.47 \\
      & 100 & 13.35 & 46.25 & 12.57 & 67.78 \\
      & 200 & 11.81 & 40.61 & 11.15 & 48.37 \\
      & 400 & 11.04 & 37.81 & 10.16 & 43.15 \\
      \cdashline{2-6}
      & Multilateration & 28.23 & 123.69 & 40.02 & 149.80 \\
    \bottomrule
  \end{tabular}
\end{table}



\subsection{EKF Benchmarks}
In addition to using \gls{gps} regression for position estimation from the OWP data, we also fused the IMU and OWP data using an \gls{ekf}. The EKF outputs both the vehicle’s coordinates and its heading angle. Specifically, the \gls{ekf} state vector is defined as 
$
\vect{x} = [x,y,v_x,v_y,\theta,a_x,a_y,b_{ax},b_{ay}]^T
$,
where $x$ and $y$ denote the 2D position, $v_x$ and $v_y$ are the velocities in the global frame along the axes $x_\mathcal{I}$ and $y_\mathcal{I}$, $\theta$ is the heading angle, $a_x$ and $a_y$ are the accelerations along $x_\mathcal{I}$ and $y_\mathcal{I}$, and $b_{ax}$, $b_{ay}$ are the accelerometer bias terms.
In our approach, the OWP data updates the position at \SI{27}{\Hz} via a pre-trained GP model, while the IMU provides \SI{200}{\Hz} updates for both position and heading. The GP model was trained with 400 samples (see Fig.~\ref{fig:GPs_results}). We also used the IMU acceleration to detect when the vehicle was stationary and applied \gls{zupt} to eliminate accumulated acceleration errors.

Tab.~\ref{tab:ekf_results} summarizes the EKF performance. It lists the P50 and P99 errors for EKF-based position estimation, the corresponding \gls{owp} errors, and the EKF heading angle errors. We used data at low, medium and high speeds, with and without the obstacle. They show that the EKF fusion approach can reduce the outliers error. For all speeds,  and with or without obstacle, the \gls{ekf} can effectively decrease the P99 errors from approximately \SI{40}{\cm} (only based on OWP) to \SI{25}{\cm}. Besides, for P50, it also improve the localization accuracy from \SI{10}{\cm} to about \SI{8}{\cm}. This shows that combining the IMU with the OWP reduces outliers and improves average accuracy. We also track the yaw angle with a mean absolute error of about \SI{8}{\degree} over more than \SI{10}{\min}.

\renewcommand{\arraystretch}{1.5}
\begin{table}[!h]
  \centering
  \caption{EKF and OWP Localization Performance. The EKF position and OWP position denote the 2D coordinates Euclidean errors. The EKF Yaw mean absolute error (MAE) represent the orientation track ability.}
  \label{tab:ekf_results}
  \resizebox{\columnwidth}{!}{%
  \begin{tabular}{l c c c c c}
    \hline
    \multirow{2}{*}{Dataset} & \multicolumn{2}{c}{EKF Position [cm]} & \multicolumn{2}{c}{OWP Position [cm]} & \multirow{2}{*}{EKF Yaw MAE} \\
    \cline{2-5}
     & P50 & P99 & P50 & P99 &  \\ 
    \hline
    \shortstack{Low speed\\With Obstacle}      & 7.50 & 25.40 & 10.60 & 37.10 & 8.569$^\circ$ \\ \hline
    \shortstack{Low speed\\Without Obstacle}   & 6.80 & 20.00 & 9.30  & 33.80 & 8.138$^\circ$ \\ \hline
    \shortstack{Medium speed\\With Obstacle}    & 8.90 & 28.30 & 10.50 & 41.20 & 6.539$^\circ$ \\ \hline
    \shortstack{Medium speed\\Without Obstacle} & 9.10 & 26.00 & 9.7  & 39.40 & 7.465$^\circ$ \\ \hline
    \shortstack{High speed\\With Obstacle}      & 9.40 & 25.60 & 10.90 & 39.30 & 4.573$^\circ$ \\ \hline
    \shortstack{High speed\\Without Obstacle}   & 8.30 & 25.10 & 11.80 & 38.40 & 11.178$^\circ$ \\ \hline
  \end{tabular}%
  }
\end{table}

\begin{figure}[!h]
  \centering
  \includegraphics[width=\linewidth]{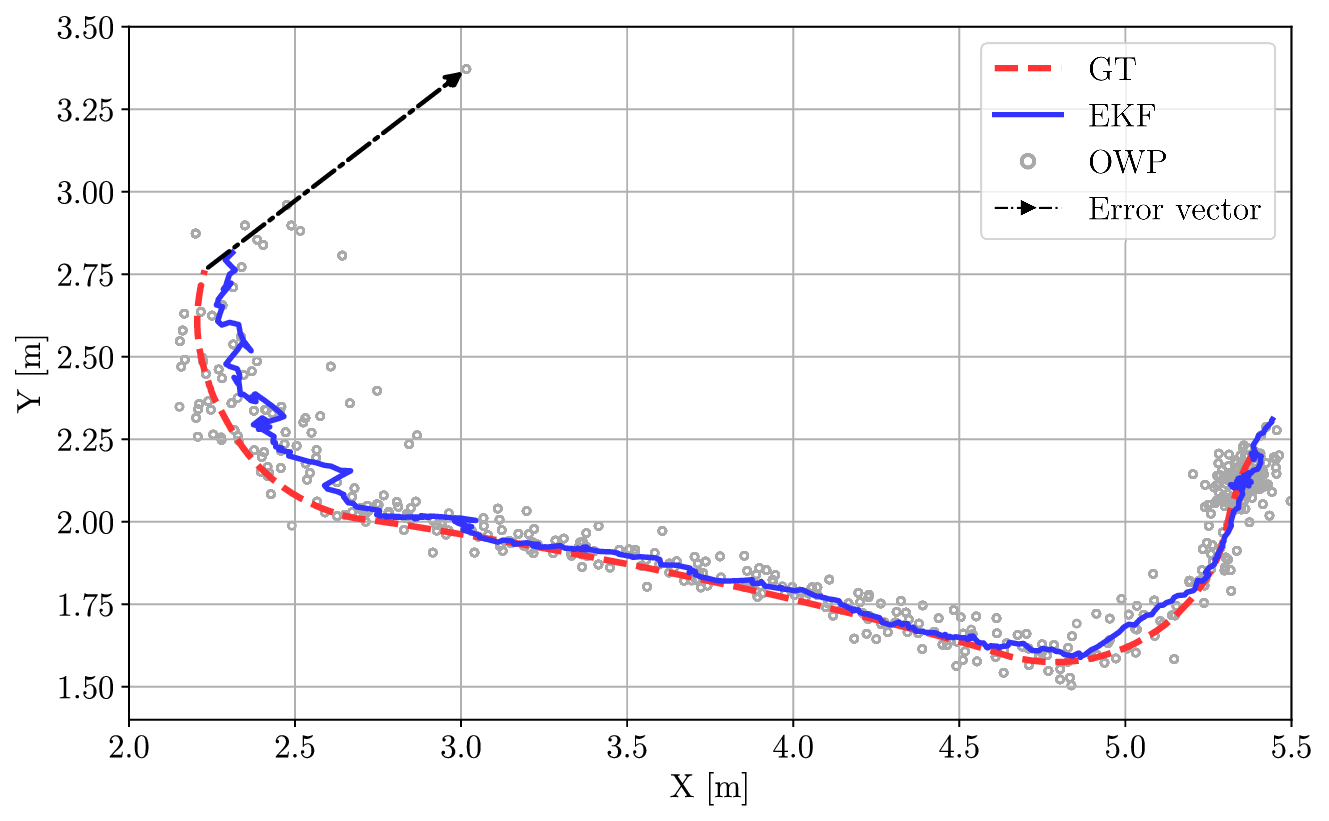}
  \caption{The GT, EKF and OWP tracked trajectory with medium speed and without obstacle.}
  \label{fig:kf_result}
\end{figure}

Fig.~\ref{fig:kf_result} display the real time \gls{ekf} results (blue line) and the OWP results (gray circles). The red line denotes the \gls{gt} measured by the Qualisys system. The black arrow marks one large error from the GT point to the OWP estimate. When the vehicle moves, random disturbances—such as shadows or vehicle vibration—could cause these outliers. This result in  one OWP estimate far from the GT, as highlighted by the arrow. The  \gls{ekf}  handles these outliers better than OWP alone.

\section{Conclusion}
\label{sec:CONCLUSION}

In this paper, we introduced the OWP‑IMU dataset, a public RSS‑based optical wireless and IMU indoor localization dataset. It contains over \SI{110}{k} samples of infrared RSS values at \SI{27}{\Hz}, IMU data at \SI{200}{\Hz}, and GT at \SI{160}{\Hz}. We recorded continuous vehicle trajectories in both LOS and NLOS conditions and at three distinct velocities. We also provided two benchmarks. The GP regression on RSS data alone reaches about \SI{10}{\cm} median error with 400 training samples. The EKF fusion of RSS and IMU cuts the P99 error from \SI{40}{\cm} down to \SI{25}{\cm} and yields about 8° mean yaw error over ten minutes of motion. By making this dataset available, we aim to help the community develop and improve localization data processing methods. Future work will explore advanced fusion algorithm (e.g., diffusion model) to further  improve accuracy and robustness.




\bibliographystyle{IEEEtran}
\bibliography{IEEEabrv,mybibfile}

\end{document}